\documentclass[prb,twocolumn,showpacs]{revtex4}   

\usepackage{graphicx}

\begin{document}

\title
{
Field-induced phase transitions in a Kondo insulator
}

\author{Takuma Ohashi}
\author{Akihisa Koga}
\author{Sei-ichiro Suga}
\author{Norio Kawakami}
\affiliation{%
Department of Applied Physics, Osaka University, 
Suita, Osaka 565-0871, Japan}%

\date{\today}

\begin{abstract}
We study the magnetic-field effect on a Kondo insulator by 
exploiting the periodic Anderson model with the Zeeman term. 
The analysis using dynamical mean field theory combined with 
quantum Monte Carlo simulations determines the detailed phase diagram 
at finite temperatures. 
At low temperatures, the magnetic field drives the Kondo insulator 
to a transverse antiferromagnetic phase, 
which further enters a polarized metallic phase at higher fields.
The antiferromagnetic transition temperature $T_c$ 
takes a maximum  when the Zeeman energy is nearly equal to
 the quasi-particle gap. 
In the paramagnetic phase above $T_c$, we find that 
the electron mass gets largest around the field
where the quasi-particle gap is closed. It is also shown that the
 induced moment of conduction electrons changes its direction
 from antiparallel to parallel to the field.
\end{abstract}
\maketitle

\section{Introduction}

There has been a continued interest in a class of compounds called 
Kondo insulators or heavy fermion semiconductors,\cite{ki} 
which develop an insulating gap at low temperatures 
via strong correlation effects. 
${\rm CeRhAs}$, ${\rm Ce_3Bi_4Pt_3}$, 
${\rm YbB_{12}}$, and ${\rm SmB_6}$ 
are well known examples, which possess small gaps of the order 
of 1-100 meV in spin and charge excitations. 
The gap formation in Kondo insulators is  
attributed to the renormalized hybridization between a broad conduction
 band and a nearly flat $f$-electron band with strong 
correlations.
In some Kondo insulators, enhanced antiferromagnetic (AF) correlations 
among localized $f$-moments dominate  the Kondo 
singlet formation, resulting in an AF long-range order.\cite{ins}
Systems located around such magnetic instability provide
hot topics in quantum critical phenomena.\cite{qcp}

The application of  a magnetic field to the Kondo insulators
also realizes 
quantum critical phenomena, which  have been 
investigated extensively. 
Experiments on ${\rm Ce_3Bi_4Pt_3},$\cite{ce} 
${\rm YbB_{12}},$\cite{yb} and ${\rm SmB_6}$\cite{sm} in high 
magnetic fields indicate closure of the Kondo-insulating gap, 
exemplifying a transition from the Kondo insulator to a 
correlated metal.  If
 the Kondo insulator is in the proximity of magnetic instability, 
 the local singlet formation gets weak, and 
the applied field may possibly trigger 
a phase transition to the AF
ordered state before it becomes a metal.

The periodic Anderson model (PAM) at half filling may be a simplified
 model to describe 
 the Kondo insulating phase\cite{hf} and the
AF  phase, depending on the $c$-$f$ hybridization 
strength $V$ and the $f$-$f$ Coulomb interaction $U$. 
The magnetic instability of the PAM has been investigated
by means of various methods such as 
slave-boson mean field theory\cite{pam_phase1,pam_ins,pam_phase2}
and dynamical mean field theory (DMFT).
\cite{jarrell,rozenberg,dmft,imai}
The effect of the magnetic field on the PAM has also been 
investigated,\cite{saso,ono,ohara,mutou1,satoh,meyer}
and the phase transition from the Kondo insulator 
to a paramagnetic metal,
which is naively expected, has been discussed. 

Recently, Beach {\it et. al.}\cite{beach} have studied 
the magnetic-field effect on the two-dimensional
Kondo lattice model using 
large-$N$  mean field theory and quantum Monte Carlo (QMC)
 simulations. Also,
Milat {\it et. al.} have presented a mean field analysis of the  PAM
in the small $U$ region and  QMC simulations of the Kondo
lattice model.\cite{milat} 
They have found that the magnetic field induces a second-order phase 
transition from the paramagnetic to the transverse  AF phase
in the  Kondo insulator. Since these works have been
concerned with the two-dimensional systems, it is desirable to
extend the investigation to three-dimensional cases. 
Also, a detailed  study including the paramagnetic phase 
at finite temperatures
is expected to provide further interesting properties in 
the Kondo insulator in a magnetic field.

In this paper, we study field-induced AF phase transitions
of the Kondo insulator in the whole temperature regime
by using the PAM at half filling.
 By exploiting DMFT combined with the QMC method,\cite{qmc} 
we show that a magnetic field induces a transverse  AF
order in the Kondo insulator. We also demonstrate that a correlated metallic
state with characteristic properties emerges in the 
paramagnetic phase slightly above the AF transition
temperature, when the Zeeman energy is nearly equal to the spin gap.

This paper is organized as follows.  In the next section, we
briefly mention the model and method, and  show the 
phase diagram obtained by DMFT. In Sec.III, we explain the 
detail of the  field-induced AF transitions, and
discuss some remarkable properties in the paramagnetic phase.
Brief summary is given in Sec. IV.

\section{Model and Phase Diagram}

Let us begin with the 
 periodic Anderson model with the Zeeman term,
\begin{eqnarray}
H &=&
	-t \sum_{\left < i,j \right >,\sigma} c_{i\sigma}^\dag c_{j\sigma}
		+ V \sum_{i,\sigma}
		\left [
			c_{i \sigma}^\dag f_{i \sigma} 
			+ \mathrm{h.c.}
		\right ] \nonumber \\
		&+& U \sum_i \left [ n^f_{i \uparrow} - 1/2 \right ]
		\left [ n^f_{i \downarrow} - 1/2 \right ] \nonumber \\
		&-& g \mu _B B \sum _i 
		\left [ S^{f}_{i,z} + S^{c}_{i,z} \right ], \nonumber
\end{eqnarray}
where $c_{i\sigma}$ ($f_{i \sigma}$) annihilates a conduction 
($f$) electron on the $i$th site with spin $\sigma$, 
$n_{i\sigma}^a = a_{i\sigma}^{\dag} a_{i\sigma}$ 
and $S^{a}_{i,z} = \frac{1}{2} ( n_{i\uparrow}^a 
- n_{i\downarrow}^a)$ with $a=c, f$. Here,
$t$ is the nearest-neighbor hopping matrix in the conduction 
band (referred to as the $c$ band in the following). We consider the 
 hyper-cubic  lattice with a bipartite property in infinite dimensions.
The bare density of states for $c$ electrons is Gaussian 
with the width $t^*$: $\rho_0(\varepsilon)
=\exp[-(\varepsilon / t^*)^2]/\sqrt{\pi t^{*2}}$.
The magnetic field $B$ applied along the $z$ direction is
coupled to both of the $c$ and $f$ electrons.  We assume that
the $g$ factors of the $c$ and $f$ electrons are
the same, and set $g \mu_B = 1$.
These simplifications would not qualitatively 
change our conclusions.

DMFT is a powerful framework to study 
strongly correlated electron systems, {\it e.g.},
the single-band Hubbard model,%
\cite{georges,jarrell_shm,zhang,%
caffarel,fisher,kajueter,bulla} %
the two-band Hubbard model,%
\cite{rozenberg_thm,held,han,maier,momoi,%
imai2,oudovenko,florens,koga} %
the PAM,%
\cite{jarrell,rozenberg,imai,saso,mutou1,meyer,%
tahvildar,pruschke,shimizu,%
vidhyadhiraja,ryouta,mutou2} etc. 
This method is justified in the limit of 
large spatial dimensions and  gives a rather good 
approximation even in three dimensions. 
In DMFT, the proper one-particle self-energy, which is independent 
of the momentum, is obtained via an effective impurity model 
embedded in a self-consistently determined medium. 
The DMFT treatment of the PAM is summarized in Appendix A.
We solve the effective impurity model using the QMC method with 
the Trotter time slices, 
$\Delta \tau = \beta/L \le 0.25$ ($\beta=1/T$). 
The number of QMC sweeps is $3.2 \times 10^6$ in each DMFT 
iteration loop.  We use the typical parameters 
$U/t^*=2.0$ and $V/t^*=0.6$ in the
following  calculation. 
For these parameters, the ground state at zero field 
 is a paramagnetic Kondo insulator, as shown 
by Jarrell {\it et. al.}\cite{jarrell} It is convenient to
 express the energy normalized by the zero field value of 
quasi-particle gap, $\Delta_0 \sim 0.145 t^*$, 
which is estimated 
from the $f$-electron spectral function at $T/t^* = 1/30$ 
(see Fig. \ref{fig3}). 

\begin{figure}[bt]
\begin{center}
\includegraphics[clip,trim=1cm 14.5cm 1cm 0cm,width=8cm]{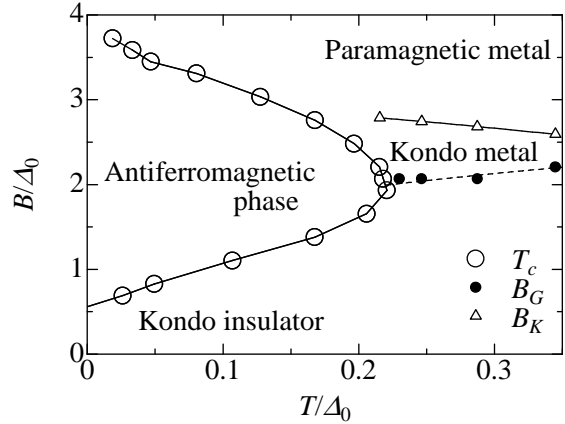}
\end{center}
\caption{Phase diagram of the PAM on the hypercubic lattice in 
infinite dimensions. The parameters are chosen as $U=2.0$ and $V=0.6$. 
The second-order AF transition temperature $T_c$
is indicated by open circles. Two  fields $B_G$ and $B_K$ 
characterize the crossover behavior related to the gap closing and
the change in the magnetization process, respectively. The region 
denoted as a Kondo metal possesses interesting physical properties 
different from ordinary heavy electrons, see text.
}
\label{fig1}
\end{figure}

We first give an overview  of the phase diagram obtained 
in this paper, which is shown in Fig.\ref{fig1}.  
At weak fields, the system is in 
the Kondo insulating phase characterized by the
gap formation of quasi-particle states, which is smoothly
connected  to the high-temperature paramagnetic phase. 
Beyond a certain critical field, there is a
second-order phase transition from  the 
paramagnetic phase to the low-temperature 
AF phase with the decrease of temperature. 
The AF phase possesses a long-range order in the  $xy$ plane,
namely, the transverse AF order. 

It is seen that the transition temperature $T_c$ takes a maximum, when the 
Zeeman energy is nearly equal to the spin gap 
$B \sim 2\Delta_0$. 
 Around this field, we find remarkable
properties in the paramagnetic phase just above $T_c$: 
they are characterized by two crossover fields, $B_G$ and $B_K$.
The quasi-particle gap is closed for fields higher
than $B_G$. At $B_K$, the magnetization of the 
$c$ electrons changes its sign. 
The region surrounded by $T_c$, $B_G$ and $B_K$ may be
referred to as a `Kondo metal', 
where the quasi-particle gap  is closed but the Kondo 
correlations are still dominant to produce the 
strong renormalization effects accompanied by the large 
mass enhancement. 
 As a result, this Kondo-metal region
exhibits interesting  properties coming from the competition 
of the Kondo correlations, the AF instability, etc.

\section{ Detail of the Results}

We now  explain how the above phase diagram is obtained in the 
DMFT framework.  We first investigate the field-induced AF transitions,
and then discuss some characteristic properties in the 
paramagnetic phase.

\subsection{Phase transitions}

An AF transition is signaled by the divergence 
of the staggered spin susceptibility $\chi_{xx}(\mathbf{Q})$
with $\mathbf{Q}=[\pi,\pi,\ldots]$, where the suffix $x$ denotes the 
direction perpendicular to the field. 
In DMFT, we can calculate this quantity via the impurity model, since 
the irreducible vertex function as well as the self-energy is 
independent of the momentum in infinite dimensions.\cite{jarrell}
The detail of the calculation is summarized in Appendix B.
 
\begin{figure}[bt]
\begin{center}
\includegraphics[clip,trim=1cm 11cm 1cm 2cm,width=8cm]{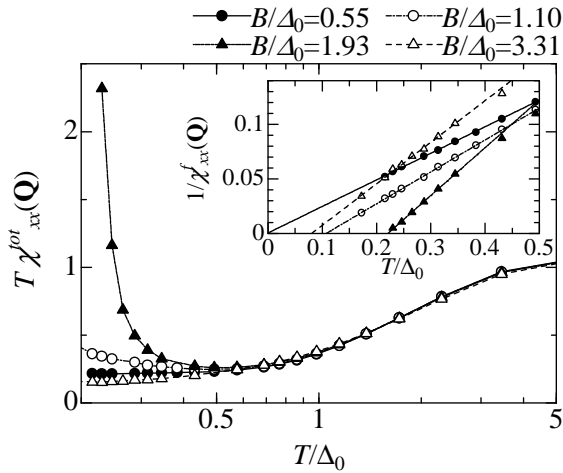}
\end{center}
\caption{Temperature dependence of the total staggered
susceptibility $\chi ^{tot}_{xx}(\mathbf{Q})$ 
with $\mathbf{Q}=[\pi,\pi,\ldots]$. 
Inset  shows the inverse of the susceptibility for $f$ electrons 
$\chi ^f_{xx}(\mathbf{Q})$. Both of
$\chi ^{tot}_{xx}(\mathbf{Q})$ and $\chi ^{f}_{xx}(\mathbf{Q})$ 
diverge at the same critical temperature. 
$\chi ^f_{xx}(\mathbf{Q})$ is well fitted with the
 form $1 / \chi ^f_{xx}(\mathbf{Q}) \propto T-T_c$, 
which enables us to
determine the transition temperature $T_c$  by a proper extrapolation 
procedure.}
\label{fig2}
\end{figure}

In Fig. \ref{fig2},  the staggered spin 
susceptibility $T \chi^{tot} _{xx}(\mathbf{Q})$ is
plotted as a function of temperature for different magnetic fields. 
In the DMFT framework, the susceptibility exhibits
 mean-field behavior, $\chi_{xx}(\mathbf{Q}) \propto [T-T_c]^{-1}$.
Thus, we evaluate the transition temperature rather accurately by extrapolating
 the inverse susceptibility to zero as a function of $T$, as 
shown in the inset of  Fig. \ref{fig2}. 
The transition temperature $T_c$ thus obtained for various values of $B$ 
is plotted in the phase diagram shown in Fig. \ref{fig1}. 
The magnetic field triggers a phase transition 
 from the paramagnetic Kondo insulator to the transverse AF
ordered state. 
The critical field between the Kondo insulator and the AF
insulator is estimated as $B \sim 0.56 \Delta_0$ at $T=0$.
In our calculation, it is somewhat difficult to precisely determine
 the  higher critical field separating the AF phase and
 the paramagnetic metallic phase at $T=0$, since the signal of 
antiferromagnetism appears 
at very low temperature with increasing $B/\Delta_0$. 
Milat {\it et. al.} concluded that the AF state should
persist up to the fully polarized state 
because of perfect nesting.\cite{milat}
Our numerical results support their conclusion,
although such a transition would occur at extremely low temperatures 
in the high field regime.  It is instructive to note that
the transition temperature $T_c$ takes a maximum 
value $T_c \sim 0.22 \Delta_0$ around the field 
$B \sim 1.93 \Delta_0$, which is almost 
the same as the spectral gap, $2\Delta_0$. 
We will see that this value of field is close to
two typical crossover fields discussed below.

\subsection{Heavy-fermion behavior in paramagnetic phase}

\begin{figure}[bt]
\begin{center}
\includegraphics[clip,trim=0cm 0.5cm 1cm 5cm,width=8cm]{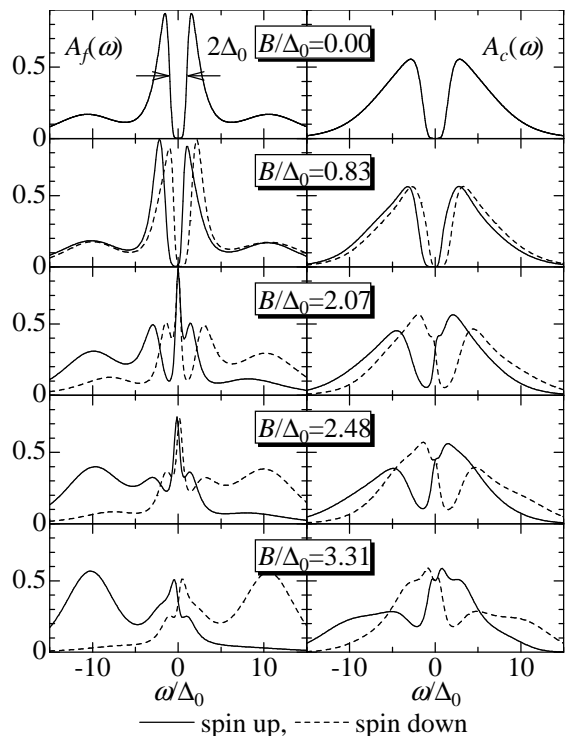}
\end{center}
\caption{Spectral functions of
$f$ (left) and $c$ (right) electrons  at $T = 0.23 \Delta_0$ 
for different values of $B/\Delta_0$. 
The energy scale $\Delta_0$ is determined by the distance 
between zero frequency and the frequency which takes the half
 value of the first peak in the $f$-electron spectral 
function at zero field. 
}
\label{fig3}
\end{figure}

We obtain dynamical quantities by applying 
the maximum entropy method (MEM) to
 the imaginary-time QMC data computed with DMFT.
In Fig. \ref{fig3}, we show the one-particle spectral functions
 for the $c$ and $f$ electrons,
$A_{c\uparrow,\downarrow}(\omega)$ and $A_{f\uparrow,\downarrow}(\omega)$, 
at $T = 0.23 \Delta_0$ for different magnetic fields. Note that
the system is in the paramagnetic phase for any magnetic fields at
this temperature.
One can see in this figure how the magnetic filed closes the 
quasi-particle gap 
and  then induces correlated metallic states.  
At $B=0$, the well-defined insulating gap exists both for
$f$ and $c$ electrons. As the field is increased
($B/\Delta_0=0.83$), the peak structure existing
beside the gap is slightly shifted in the 
presence of the Zeeman splitting. 
 For $B/\Delta_0  \sim 2$, 
the spectral gap almost 
disappears.  A remarkable point in this case is that 
$f$ electrons get
 another sharp peak structure around  $\omega =0$, as seen for 
the third panel in Fig. \ref{fig3}.
This implies that 
strong Kondo correlations still persist even in such 
fields,  driving the system  to a correlated metallic 
state (Kondo metal in Fig. \ref{fig1})
immediately after  the quasi-particle gap is closed by the field.
If the field is further
increased, such a sharp quasi-particle peak disappears, 
and the system gradually changes to 
an ordinary paramagnetic metal with the spin polarization.

\begin{figure}[bt]
\begin{center}
\includegraphics[clip,trim=1cm 6cm 1cm 3cm,width=8cm]{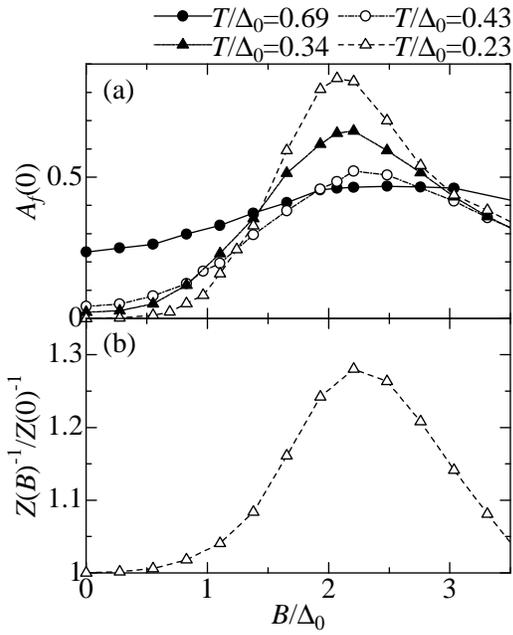}
\end{center}
\caption{(a) The $f$-electron spectral function at the Fermi level 
$A_f(0)$ as a function of $B/\Delta_0$, which is used 
to determine $B_G$; (b) the derivative of the $f$-electron
self-energy  $Z(B)^{-1}$ at 
$T=0.23 \Delta_0$ normalized by its zero-field 
value $Z(0)^{-1} (\simeq 2)$. Here 
 $Z$ is computed numerically by the formula
 $Z={\mathrm I \mathrm m} \Sigma(\omega_0)/\omega_0$ 
with $\omega_0=\pi T$.
}
\label{fig4}
\end{figure}

In order to see the above characteristics  
clearly, we show the $f$-electron spectral weight at the
Fermi level, $A_f(0)=[A_{f_\uparrow}(0)+A_{f_\downarrow}(0)]/2$, 
in Fig. \ref{fig4} (a) 
as a function of the magnetic field at different temperatures. 
It may be legitimate to define the crossover field $B_G$
 characterizing the gap-closing
by the field where $A_f(0)$ takes its maximum as a function of $B$.
The crossover field  thus estimated, $B_G  \simeq 2 \Delta_0$, 
 is consistent with the known 
results that the spin gap is nearly equal
to the quasi-particle gap in infinite dimensions.\cite{jarrell,mutou2} 
It is seen that $B_G$ is almost independent of the temperature
in the range shown in Fig.\ref{fig1}. As a reference, 
in Fig. \ref{fig4} (b) we show the numerical derivative of 
the $f$-electron 
self-energy, $Z$, which would correspond to the inverse  of the
mass-enhancement factor at low temperatures. 
It is seen that $Z^{-1}$ indeed takes its maximum 
around the field $B_G$, which seems consistent
 with the development of the 
sharp quasi-particle peak in Fig. \ref{fig3}.
However, one notices that the mass enhancement (i.e. development 
of the sharp peak) observed 
around $B/\Delta_0  \sim 2$ in Fig. \ref{fig3}
is larger than that in Fig. \ref{fig4}(b).
This difference comes from our numerical differentiation
 procedure with discrete frequency
in Fig. \ref{fig4}(b), where we have naively 
used the formula in the 
Matsubara form, $Z={\mathrm I \mathrm m} \Sigma(\omega_0)/\omega_0$
with  $\omega_0=\pi T$.  This formula may not pick up
the sharp structure of the spectrum around $B/\Delta_0 \sim 2$ 
in Fig. \ref{fig3}.
Therefore, the actual mass enhancement 
should be  somewhat larger than that displayed in 
Fig. \ref{fig4}(b).

%

\begin{figure}[bt]
\begin{center}
\includegraphics[clip,trim=1cm 0.5cm 0cm 5cm,width=8cm]{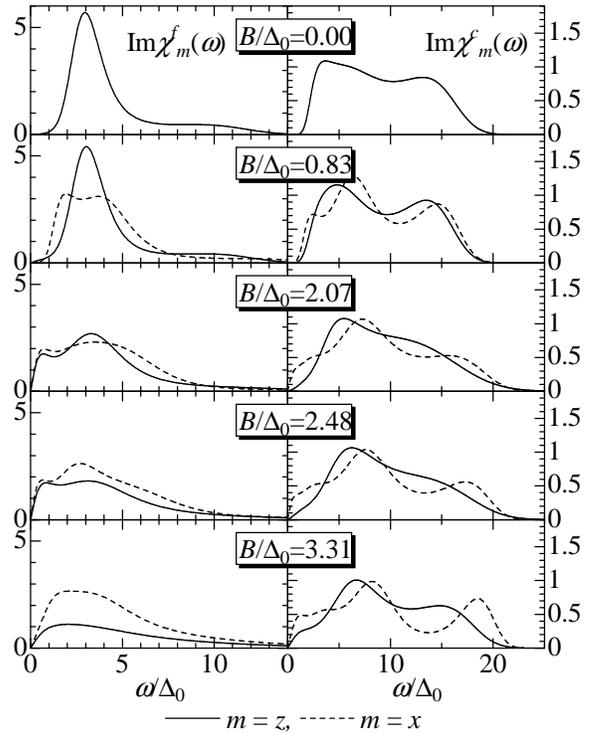}
\end{center}
\caption{
Dynamical spin-correlation function, $\mathrm{Im}\chi^{m}_{\alpha} (\omega)$, 
for $f$ (left panel) and $c$ (right panel) electrons 
at $T/\Delta_0 = 0.23 $ for 
different values of $B/\Delta_0$. Solid (broken) lines represent
the $z$- ($x$-) component of the  spin correlation function, where 
the $z$ axis is parallel to the applied field.
}
\label{fig5}
\end{figure}

We further calculate the dynamical spin-correlation function to 
directly observe the magnetic-field  effects on spin excitations. 
The imaginary-time  spin-correlation function 
$\chi^{m}_{\alpha} (\tau)$, 
defined as
$
\chi^{m}_{\alpha} (\tau) = \left \langle
\mathrm{T}_\tau 
S_{i,\alpha}^{m} \left( \tau \right ) 
S_{i,\alpha}^{m} \left ( 0 \right ) 
 \right \rangle
$
with $m = c,f$ and $\alpha = z,x$, is calculated by the QMC method and is 
analytically continued to the real frequency with MEM. 
In Fig. \ref{fig5}, we show the field dependence of 
$\mathrm{Im} \chi^{m}_{\alpha} (\omega)$ 
at a given temperature, $T/\Delta_0=0.23$, slightly larger than the 
maximum AF transition temperature.

At zero field,
there are  two characteristic features in the dynamical 
spin-correlation functions.
One is a low-energy peak (shoulder) structure
in $\mathrm{Im}\chi_\alpha^f(\omega)$ 
($\mathrm{Im}\chi_\alpha^c(\omega)$)
due to spin-triplet excitations,
which determine the size of the spin gap.
The other is a broad continuum in the high-energy region 
due to the dispersion of $c$ electrons, which features a higher-energy
hump in $\mathrm{Im}\chi_\alpha^c(\omega)$.
When a magnetic field is introduced,  
the spin-triplet excitations 
split into three distinct excitations specified by $S_z^{tot}$.
At $B/\Delta_0=0.83$, 
the magnetic field little affects the shape of
$\mathrm{Im}\chi_z^{f,c}(\omega)$,
whereas it splits the low-energy peak of 
$\mathrm{Im}\chi_x^{f,c}(\omega)$ into
two sub-peaks.  The results are naturally understood, because
the triplet excitations observed in 
$\mathrm{Im}\chi_z^{f,c}(\omega)$ ($\mathrm{Im}\chi_x^{f,c}(\omega)$)
are those with $S_z^{tot}=0$ ($S_z^{tot}=\pm 1$):
the latter causes the splitting in the spin excitation spectrum.
If the field is further increased,  the spin gap
disappears around $B/\Delta_0 \sim 2$,
making the Kondo singlet state unstable. Note that around this 
field, $\mathrm{Im}\chi_z^f(\omega)$ develops another hump structure 
around $\omega=0$, which is consistent with the fact that 
heavy-fermion  states with  the enhanced mass
appear in the one-particle spectrum (Fig. \ref{fig3}).
 Another noticeable point is that  $\mathrm{Im}\chi_z^c(\omega)$
does not increase its weight around $\omega=0$, although
$\mathrm{Im}\chi_x^c(\omega)$ as 
well as $\mathrm{Im}\chi_{z,x}^f(\omega)$
already has a reasonable weight (see data at 
$B/\Delta_0=2.07$). Furthermore, $\mathrm{Im}\chi_z^c(\omega)$ 
eventually develops a shoulder-like structure around $\omega=0$
at $B/\Delta_0=3.31$, 
where the heavy-fermion behavior in $\mathrm{Im}\chi_z^f(\omega)$ is 
already suppressed. In this way, the $z$-component spin excitations
 of $c$ electrons exhibit unusual behavior around the 
field $B/\Delta_0 = 2 \sim 3$.

\begin{figure}[bt]
\begin{center}
\includegraphics[clip,trim=1cm 6cm 1cm 3cm,width=8cm]{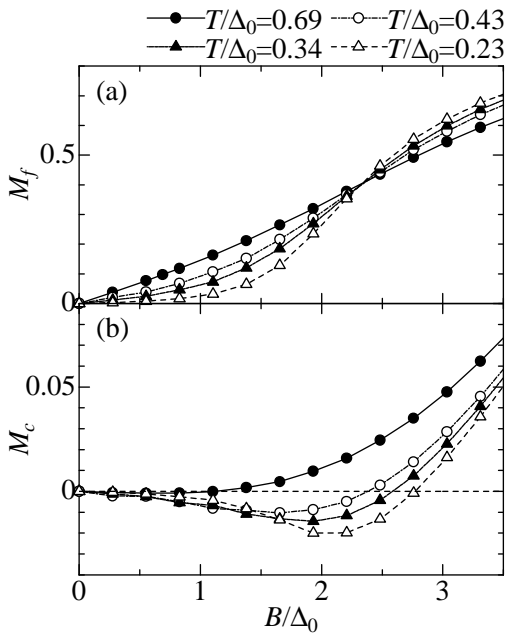}
\end{center}
\caption{
The magnetization process of $f$ (a) and $c$ (b)
electrons. 
$M_f$ and $M_c$ are defined by 
$M_f=\langle n_{f\uparrow}-n_{f\downarrow} \rangle$ and 
$M_c=\langle n_{c\uparrow}-n_{c\downarrow} \rangle$.
}
\label{fig6}
\end{figure}

To make the above point clear, we turn to another 
interesting crossover behavior 
denoted by the line $B_K$ (see Fig.\ref{fig1}) 
in the paramagnetic phase above $T_c$. This line is 
determined by the change in the magnetization process
for $c$ electrons.  Shown in Figs. \ref{fig6} (a) and (b) 
are the magnetization of $f$ and $c$ electrons 
at several different temperatures.
It is found that 
the magnetization of $f$ electrons is always positive, 
while that of $c$ electrons is negative for weak fields and 
gets positive beyond a certain magnetic field. 
We define $B_K$ by the magnetic field at which the 
magnetization of $c$ electrons changes its sign.  
This field $B_K$ 
roughly corresponds to the strength of the local singlet formation. 
At weak fields,
the Kondo effect is still dominant to align the spins 
of $c$ and $f$ electrons in the opposite directions,
which may give rise to the negative magnetization in 
$c$ electrons. 
Such nonmonotonic behavior is also seen in the spectral 
function of $c$ electrons in Fig. \ref{fig3}. 
The spectral function of $c$ electrons with up spin is 
shifted to the low-energy side, while its weight 
is considerably suppressed. These
results are also consistent with what we have observed in the 
small-$\omega$ behavior  of
 $\mathrm{Im}\chi^{x}_{c} (\omega)$ for  
$B/\Delta_0 =2  \sim 3$  in Fig. \ref{fig5}.

Summarizing the results found in the normal phase,
we can say that around the field where $T_c$ takes a maximum,
various correlation effects such as the Kondo-gap formation, 
the AF instability, etc. compete with each other, 
yielding characteristic properties in correlated electrons. 
Such features are different from ordinary heavy 
fermions without magnetic fields.

\section{Summary} 

We have investigated the effects of the magnetic field 
on  the Kondo insulator by 
applying DMFT to the PAM at finite temperatures. 
There are several characteristic features in the phase diagram.
At low temperatures, we have observed the quantum 
phase transition from the Kondo insulator to the transverse AF phase,
in accordance with the results obtained in the 
two-dimensional case. 
 We have also found some remarkable properties 
in the paramagnetic regime. Namely,
in the paramagnetic phase close to the maximum
AF transition temperature, heavy-fermion states with the enhanced 
mass are realized, where the competition of the 
magnetic field and the Kondo correlations results in 
characteristic features in the one-particle spectrum,
the dynamical spin susceptibility and 
the magnetization process.  
It may be interesting to experimentally
investigate static as well as dynamical properties 
in this region.
Such investigation may provide a unique
example of field-induced heavy-fermion states
without phase transitions.

\acknowledgements

The authors thank D. S. Hirashima,  A. 
Oguri and T. Saso for valuable discussions.  
A part of numerical computations was done at the Supercomputer Center 
at the Institute for Solid State Physics, University of Tokyo. 
This work was partly supported by a Grant-in-Aid from the Ministry 
of Education, Science, Sports and Culture of Japan.

\appendix
\section{Formulation of DMFT}
In this Appendix, we summarize the DMFT formulation of  the 
PAM.\cite{dmft,jarrell,rozenberg}
In DMFT, the original lattice model is mapped onto 
an effective impunity model 
with the effective medium determined self-consistently. 
This medium is described in terms of  a cavity Green's function 
${\bf \cal G}(i \omega _n)$, which represents the motion of 
correlated electrons
on all sites except an impurity site. 
The local Green's function ${\bf G}(i \omega _n)$ 
is then expressed in terms of the momentum-independent self-energy 
$\Sigma (i \omega _n)$ 
and the cavity Green's function ${\cal G}(i \omega _n)$ as 
\begin{equation}
{\bf G}^{\sigma}(i \omega _n)=\left [ 
  {\bf \cal G}^{\sigma} (i \omega _n)^{-1} 
  - {\bf \Sigma} ^{\sigma} (i \omega _n) \right ] ^{-1},
\label{impurity}
\end{equation}
where the Green's functions and the self-energy are expressed
in the matrix form, 
\begin{eqnarray}
{\bf G}^{\sigma}(i \omega _n)&=&\left [
  \begin{array}{cc}
  G_{cc}^{\sigma}(i \omega _n) & G_{cf}^{\sigma}(i \omega _n) \\
  G_{fc}^{\sigma}(i \omega _n) & G_{ff}^{\sigma}(i \omega _n)
  \end{array} 
\right ], \\
{\bf \cal G}^{\sigma}(i \omega _n)&=&\left [
  \begin{array}{cc}
  {\cal G}_{cc}^{\sigma}(i \omega _n) 
    & {\cal G}_{cf}^{\sigma}(i \omega _n) \\
  {\cal G}_{fc}^{\sigma}(i \omega _n) 
    & {\cal G}_{ff}^{\sigma}(i \omega _n)
  \end{array}
\right ], \\
{\bf \Sigma} ^{\sigma} (i \omega _n)&=&\left [
  \begin{array}{cc}
  0 & 0 \\
  0 & \Sigma ^{\sigma}_{f}(i \omega _n)
  \end{array}
\right ].
\end{eqnarray}

On the other hand, the local Green's function $G (i\omega_n)$
can be written as
\begin{eqnarray}
G^\sigma(i\omega_n)&=& \int d \varepsilon \frac{\rho_0 ( \varepsilon )}
{\xi^\sigma (i\omega_n,\varepsilon)}
\left [
  \begin{array}{cc}
  \xi _f^\sigma (i\omega_n) & V \\
  V & \xi _c^\sigma (i\omega_n, \varepsilon ) 
  \end{array}
\right ] \nonumber \\
\label{lattice}
\end{eqnarray}
with 
\begin{eqnarray}
\xi_c^\sigma \left ( i\omega_n, \varepsilon \right ) 
  &=& i\omega_n + \frac{1}{2} \sigma g \mu _B B - \varepsilon , \\
\xi_f^\sigma \left ( i\omega_n \right ) 
  &=& i\omega_n - \frac{U}{2} + \frac{1}{2} \sigma g \mu _B B 
      - \Sigma_f \left ( i\omega_n \right ), \\ 
\xi^\sigma \left ( i\omega_n, \varepsilon \right ) 
  &=& \xi_c^\sigma \left ( i\omega_n, \varepsilon \right ) 
      \xi_f^\sigma \left( i\omega_n \right ) - V^2.
\end{eqnarray}
By solving equations (\ref{impurity}) and (\ref{lattice}) self-consistently,
we can discuss the effects of electron correlations on 
local properties in the PAM.
In this paper, we obtain the imaginary-time Green's function 
$G^\sigma (\tau)$ by solving the impurity problem 
with the QMC method.\cite{qmc}

\section{Magnetic Susceptibility}

To discuss the instability to an AF ordered phase, 
it is necessary to obtain the momentum-dependent susceptibility 
$\chi_{xx}(\mathbf{q})$ in the PAM. 
Here, we explain how  this quantity is formulated
in the framework of DMFT.\cite{dmft,jarrell}

We first consider the local two-particle propagator 
in the effective impurity system, 
\begin{eqnarray}
\lefteqn{
\chi_{+- i}^{f} \left ( \tau_1,\tau_2,\tau_3,\tau_4 \right )
}\hspace{1cm} \nonumber \\
  &=& \left \langle \mathrm{T}_\tau 
    f^\dag_{i \uparrow} \left ( \tau_1 \right )
    f_{i \downarrow} \left ( \tau_2 \right )
    f^\dag_{i \downarrow} \left ( \tau_3 \right )
    f_{i \uparrow} \left ( \tau_4 \right )
    \right \rangle, \\
\lefteqn{
\chi_{-+ i}^{f} \left ( \tau_1,\tau_2,\tau_3,\tau_4 \right )
} \hspace{1cm} \nonumber \\
  &=& \left \langle \mathrm{T}_\tau 
    f^\dag_{i \downarrow} \left ( \tau_1 \right )
    f_{i \uparrow} \left ( \tau_2 \right )
    f^\dag_{i \uparrow} \left ( \tau_3 \right )
    f_{i \downarrow} \left ( \tau_4 \right )
    \right \rangle .
\end{eqnarray}
We obtain the vertex function 
$\Gamma (i\nu_l=0;i\omega_m,i\omega_n)$
by solving the Bethe-Salpeter equation,
\begin{equation}
\chi^f_i = \chi^{f0}_i + \chi^{f0}_i \Gamma \chi^f_i,
\end{equation}
where $\chi^{f0}_i$ is the non-interacting part of the 
local two-particle propagator, 
\begin{equation}
\chi^{f0}_{i} \left ( i\nu_l=0;i\omega_m,i\omega_n \right )
  = - \frac{\delta_{m,n}}{\beta}
    G_{ff}^{\uparrow} \left( i \omega _n \right )
    G_{ff}^{\downarrow} \left( i \omega _n \right ).
\end{equation}
By using the vertex function $\Gamma$,
we obtain the two-particle propagator for correlated $f$-electrons 
in the lattice system,
\begin{equation}
\chi^f_{\mathbf{q}} = \chi^{f0}_{\mathbf{q}} 
+ \chi^{f0}_{\mathbf{q}} \Gamma \chi^f_{\mathbf{q}},
\end{equation}
where $\chi^{f0}_{\mathbf{q}}$ is the bare
two-particle propagator in the lattice system given by, 
\begin{eqnarray}
\lefteqn{
\chi^{f0}_{+- {\mathbf{q}}} \left 
  ( i\nu_l=0;i\omega_m,i\omega_n \right )
  } \hspace{7cm} \nonumber \\
  = -\frac{\delta_{n,m}}{N\beta} \sum_{\mathbf{k}} 
    G^{ff}_{\mathbf{k}\uparrow} \left ( i\omega_n \right )
    G^{ff}_{\mathbf{k}+\mathbf{q}\downarrow} \left ( i\omega_n \right ), 
\label{app1} \\
\lefteqn{
\chi^{f0}_{-+ \mathbf{q}} \left ( i\nu_l=0;i\omega_m,i\omega_n \right )
} \hspace{7cm} \nonumber \\
  = -\frac{\delta_{n,m}}{N\beta} \sum_{\mathbf{k}} 
    G^{ff}_{\mathbf{k}\downarrow} \left ( i\omega_n \right )
    G^{ff}_{\mathbf{k}+\mathbf{q}\uparrow} \left ( i\omega_n \right ).
\label{app2}
\end{eqnarray}
Equations (\ref{app1}) and (\ref{app2}) can be rewritten as
\begin{eqnarray}
\lefteqn{
\chi_{+- \mathbf{q}}^{f0}(i\nu_l=0;i\omega_m,i\omega_n) =
}  \nonumber \\
&& - \frac{\delta_{m,n}}{\beta} \int dx_1 dx_2 
     \Delta_\mathbf{q}(x_1,x_2)
     \frac{\xi_c^{\uparrow}(i\omega_m,x_1)
           \xi_c^{\downarrow}(i\omega_m,x_2)}
          {\xi^{\uparrow}(i\omega_m,x_1)
           \xi^{\downarrow} (i\omega_m,x_2)}, \nonumber \\ \\
\lefteqn{
\chi_{-+ \mathbf{q}}^{f0}(i\nu_l=0;i\omega_m,i\omega_n) =
}  \nonumber \\
&& - \frac{\delta_{m,n}}{\beta} \int dx_1 dx_2 
     \Delta_\mathbf{q}(x_1,x_2)
     \frac{\xi_c^{\downarrow}(i\omega_m,x_1)
           \xi_c^{\uparrow}(i\omega_m,x_2)}
          {\xi^{\downarrow}(i\omega_m,x_1)
           \xi^{\uparrow}(i\omega_m,x_2)}, \nonumber \\
\end{eqnarray}
with the two-particle density of states $\Delta_q(x_1,x_2)$ 
defined as 
\begin{equation}
\Delta_\mathbf{q}(x_1,x_2) = \frac{1}{N} {\sum_\mathbf{k}}
  \delta(x_1-\varepsilon_{\mathbf{k}}) 
  \delta(x_2-\varepsilon_{\mathbf{k}+\mathbf{q}}). 
\end{equation}
If we focus on the center or the corner of the Brillouin zone, 
this function is reduced to the simpler form,
\begin{eqnarray}
\Delta_{\mathbf{q}}(x_1,x_2) = \left \{
\begin{array}{cl}
  \delta (x_1-x_2) \rho_0(x_1), &
  \mathbf{q}=\left [ 0,0,\ldots \right] \\
  \delta (x_1+x_2) \rho_0(x_1), &
  \mathbf{q}=\left [ \pi,\pi,\ldots \right]
\end{array}
\right. . \nonumber \\
\end{eqnarray}
Using the above equations, we can calculate the momentum-dependent
two-particle propagator 
$\chi_{\mathbf{q}}^{f}(i\nu_l=0;i\omega_m,i\omega_n)$. 
We then obtain the static susceptibility,
\begin{equation}
\chi^f(\mathbf{q}) = \frac{1}{\beta} \sum_{m,n} 
  \chi^f_{\mathbf{q}} (i\nu_l=0;i\omega_m,i\omega_n). 
\end{equation}

The total two-particle propagator in the lattice system, 
defined as 
$
\chi_\mathbf{q}^{tot} = \chi_\mathbf{q}^{c} 
+ \chi_\mathbf{q}^{cf} + \chi_\mathbf{q}^{fc}
+\chi_\mathbf{q}^{f},
$
can be expressed by that for correlated 
$f$-electrons $\chi^f_{\mathbf{q}}$ as
\begin{eqnarray}
\chi_{\mathbf{q}}^{tot} &=& \chi_{\mathbf{q}}^{c0} 
  - \chi_{\mathbf{q}}^{cf0} 
  \left ( \chi_{\mathbf{q}}^{f0} \right ) ^{-1} 
  \chi_{\mathbf{q}}^{fc0} \nonumber \\
  &+& \left[ {\mathbf I} + \chi_{\mathbf{q}}^{cf0} 
  \left ( \chi_{\mathbf{q}}^{f0} \right ) ^{-1} \right ] 
  \chi_{\mathbf{q}}^{f} 
  \left [ {\mathbf I} + 
  \left ( \chi_{\mathbf{q}}^{f0} \right ) ^{-1} 
  \chi_{\mathbf{q}}^{fc0} \right ],
\nonumber \\
\end{eqnarray}
where ${\mathbf I}$ the unit matrix, and 
$\chi_{\mathbf{q}}^{c0} (i\nu_l=0;i\omega_m,i\omega_n)$
and 
$\chi_{\mathbf{q}}^{cf0} (i\nu_l=0;i\omega_m,i\omega_n)$ 
are the $c$-$c$ and $c$-$f$ elements of the bare two-particle 
propagator, which are obtained in the same way as 
$\chi_{\mathbf{q}}^{f0} (i\nu_l=0;i\omega_m,i\omega_n)$.
We then obtain the total static-susceptibility 
$\chi^{tot}(\mathbf{q})$, 
summing over the Matsubara frequency in 
$\chi_\mathbf{q}^{tot} (i\nu_l=0;i\omega_m,i\omega_n)$.


\begin{thebibliography}{99}

\bibitem{ki} P. S. Riseborough, Adv. Phys, {\bf 49}, 257 (2000). 

\bibitem{ins} T. E. Mason, G. Aeppli, A. P. Ramirez, 
K. N. Clausen, C. Broholm, N. St\"ucheli, 
E. Bucher and T. T. M. Palstra, 
Phys. Rev. Lett. {\bf 69}, 490 (1992);  
%
H. Kadowaki, T. Sato, H. Yoshizawa, 
T. Ekino, T. Takabatake, H. Fujii, 
L. P. Regnault and Y. Isikawa, 
J. Phys. Soc. Jpn. {\bf 63}, 2074 (1994). 

\bibitem{qcp}
H. v. Lohneysen, J. Phys.: Condens. Matter {\bf 8}, 9689 (1996);
G. R. Stewart, Rev. Mod. Phys. {\bf 73}, 794 (2001); 
%
Q. Si, S. Rabello, K. Ingersent and J. L. Smith, 
Nature {\bf 413} (2000). 
%

\bibitem{ce} M. Jaime, R. Movshovich, G. R. Stewart, 
W. P. Beyermann, M. G. Berisson, M. F. Hundley, 
P. C. Canfield and J. L. Sarrao, 
Nature {\bf 405}, 160 (2000). 

\bibitem{yb} K. Sugiyama, F. Iga, M. Kasaya, 
T. Kasuya, and M. Date, 
J. Phys. Soc. Jpn. {\bf 57}, 3946 (1988). 

\bibitem{sm} J. C. Cooley, C. H. Mielke, W. L. Hults, 
J. D. Goettee, M. M. Honold, R. M. Modler, 
A. Lacerda, D. G. Rickel and J. L. Smith, 
J. Supercond. {\bf 12}, 171 (1999). 

\bibitem{hf} N. Grewe and F. Steglich, 
{\it Handbook on the Physics and Chemistry of Rare Earths}, 
edited by K. A. Gschneidner, Jr. and L. Eyring 
(North-Holland, Amsterdam, 1991), Vol. 14, p. 343. 

\bibitem{pam_phase1} V. Dorin and P. Schlottmann, 
Phys. Rev. B {\bf 46}, 10800 (1992). 

\bibitem{pam_ins} P. S. Riseborough, 
Phys. Rev. B {\bf 45}, 13984 (1992). 

\bibitem{pam_phase2} R. Doradzi\'nski and J. Spalek, 
Phys. Rev. B {\bf 58}, 3293 (1998). 

\bibitem{jarrell}
M. Jarrell, H. Akhlaghpour, T. Pruschke
Phys. Rev. Lett. {\bf 70}, 1670 (1993); 
M. Jarrell Phys. Rev. B {\bf 51}, 7429 (1995). 

\bibitem{rozenberg} M. J. Rozenberg, 
Phys. Rev. B {\bf 52}, 7369 (1995). 

\bibitem{dmft} A. Georges, G. Kotliar, W. Krauth 
and M. J. Rozenberg, Rev. Mod. Phys. {\bf 68}, 13 (1996). 

\bibitem{imai} Y. Imai and N. Kawakami, 
Acta. Phys. Pol. B {\bf 34}, 779 (2003). 

\bibitem{saso}
T. Saso, J. Phys. Soc. Jpn. {\bf 66}, 1175 (1995); 
T. Saso and M. Itoh, Phys. Rev. B {\bf 53}, 6877 (1996); 
T. Saso and H. Harima, 
J. Phys. Soc. Jpn. {\bf 72}, 1131  (2003).

\bibitem{ono}
Y. \=Ono, J. Phys. Soc. Jpn. {\bf 67}, 2197 (1998). 

\bibitem{ohara}
K. Ohara, K. Hanzawa and K. Yosida, 
J. Phys. Soc. Jpn. {\bf 68}, 521 (1999). 

\bibitem{mutou1}
T. Mutou, Phys. Rev. B {\bf 62}, 15589 (2000). 

\bibitem{satoh}
H. Satoh and F. J. Ohkawa, 
Phys. Rev. B {\bf 63}, 184401 (2001). 

\bibitem{meyer} 
D. Meyer and W. Nolting, Phys. Rev. B {\bf 64}, 52402 (2001).

\bibitem{beach} K. S. D. Beach, P. A. Lee 
and P. Monthoux, Phys. Rev. Lett. {\bf 92}, 26401 (2004). 

\bibitem{milat} I. Milat, F. Assaad and M. Sigrist, 
Eur. Phys. J. B, {\bf 38}, 571 (2004).

\bibitem{qmc} J. E. Hirsch and R. M. Fye, 
Phys. Rev. Lett. {\bf 56}, 2521 (1986). 

%
%
\bibitem{georges}
A. Georges and G. Kotliar, Phys. Rev. B {\bf 45}, 6479 (1992).
%
\bibitem{jarrell_shm}
M. Jarrell, Phys. Rev. Lett. {\bf 69}, 168 (1992).
%
\bibitem{zhang}
X. Y. Zhang, M. J. Rozenberg, and G. Kotliar, 
Phys. Rev. Lett. {\bf 70}, 1666 (1993).
%
\bibitem{caffarel}
M. Caffarel and W. Krauth, Phys. Rev. Lett. {\bf 72}, 1545 (1994). 
%
\bibitem{fisher}
D. S. Fisher, G. Kotliar, and G. Moeller, 
Phys. Rev. B {\bf 52}, 17112 (1995).
%
\bibitem{kajueter}
H. Kajueter and G. Kotliar, 
Phys. Rev. Lett. {\bf 77}, 131 (1996).
%
\bibitem{bulla} 
R. Bulla, 
Phys. Rev. Lett. {\bf 83}, 136 (1999). 
%
%
\bibitem{rozenberg_thm}
M. J. Rozenberg, Phys. Rev. B {\bf 55}, R4855 (1997). 
%
\bibitem{held}
K. Held and D. Vollhardt, Eur. Phys. J. B {\bf 5}, 473 (1998).
%
\bibitem{han}
J. E. Han, M. Jarrell, and D. L. Cox, 
Phys. Rev. B {\bf 58}, R4199 (1998). 
%
\bibitem{maier}
Th. Maier, M. B. Z\"olfl, Th. Pruschke, and J. Keller, 
Eur. Phys. J. B {\bf 19}, 377 (1999).
%
\bibitem{momoi}
T. Momoi and K. Kubo, 
Phys. Rev. B {\bf 58}, R567 (2000).
%
\bibitem{imai2}
Y. Imai and N. Kawakami, 
J. Phys. Soc. Jpn. {\bf 70}, 2365 (2001).
%
\bibitem{oudovenko}
V. S. Oudovenko and G. Kotliar, 
Phys. Rev. B {\bf 65}, 075102 (2002). 
%
\bibitem{florens}
S. Florens, A. Georges, G. Kotliar and O. Parcollet, 
Phys. Rev. B {\bf 66}, 205102 (2002).
%
\bibitem{koga}
A. Koga, Y. Imai, and N. Kawakami, 
Phys. Rev. B {\bf 66}, 165107 (2002); 
%
A. Koga, T. Ohashi, Y. Imai, 
S. Suga and N. Kawakami, 
J. Phys. Soc. Jpn. {\bf 72}, 1306 (2003); 
%
A. Koga, N. Kawakami, 
T. M. Rice and M. Sigrist, 
Phy. Rev. Lett. {\bf 92}, 216402 (2004).
%
%
\bibitem{mutou2} 
T. Mutou and D. Hirashima, 
J. Phys. Soc. Jpn. {\bf 63}, 4475(1994); 
{\bf 64}, 4799 (1995).
%
\bibitem{tahvildar}
A. N. Tahvildar-Zadeh, M. Jarrell
 and J. K. Freericks, 
Phys. Rev. B {\bf 55}, R3332 (1997); 
%
\bibitem{pruschke}
Th. Pruschke, R. Bulla, M. Jarrell, 
Phys. Rev. B {\bf 61}, 12799 (2000). 
%
\bibitem{shimizu} 
Y. Shimizu, O. Sakai and A. C. Hewson, 
J. Phys. Soc. Jpn. {\bf 69}, 1777 (2000). 

\bibitem{vidhyadhiraja}
N. S. Vidhyadhiraja, V. E. Smith, 
D. E. Logan and H. R. Krishnamurthy, 
J. Phys.: Condens. Matter {\bf 15}, 4045 (2003). 
%
\bibitem{ryouta}
R. Sato, T. Ohashi, A. Koga and N. Kawakami, 
cond-mat/0404553

\end{thebibliography}
\end{document}